# Exciton enhanced nonlinear optical responses in monolayer h-BN and $MoS_2$: Insight from first-principles exciton-state coupling formalism and calculations

Jiawei Ruan[$1,2], Yang-Hao Chan[$,3,4]*, Steven G. Louie[1,2]*

[1] *Department of Physics, University of California at Berkeley, Berkeley, California 94720, USA*

[2] *Materials Sciences Division, Lawrence Berkeley National Laboratory, Berkeley, California 94720, USA.*

[3] *Physics Division, National Center of Theoretical Physics, Taipei 10617, Taiwan*

[4] *Institute of Atomic and Molecular Sciences, Academia Sinica, Taipei 10617, Taiwan*

* yanghao@gate.sinica.edu.tw

* sglouie@berkeley.edu

[$]J. R. and Y. C. contributed equally to this paper.

**Abstract:**

Excitons are vital in the photophysics of materials, especially in low-dimensional systems. The conceptual and quantitative understanding of excitonic effects in nonlinear optical (NLO) processes is more challenging compared to linear ones. Here, we present an *ab initio* approach to second-order NLO responses, incorporating excitonic effects, that employs an exciton-state coupling formalism and allows a detailed analysis of the role of individual excitonic states. Taking monolayer h-BN and $MoS_2$ as two prototype 2D materials, we calculate their second harmonic generation (SHG) susceptibility and shift current conductivity tensor. We find strong excitonic enhancement requires that the resonant excitons are not only optically bright themselves, but also be able to couple strongly to other bright excitons. Our results explain the occurrence of two strong peaks in the SHG of monolayer h-BN and why the A and B excitons of $MoS_2$ unexpectedly exhibit minimal excitonic enhancement in both SHG and shift current generation.

In low-dimensional semiconductors, strongly correlated electron-hole pairs known as excitons (either strongly bound or in resonance with the two-particle continuum) dominate the low-energy excitations and play a key role in light-matter interactions. Understanding excitonic effects is essential for fundamental science and optoelectronic device applications. It is well established that the strong enhancement of linear optical response in low-dimensional materials is due to strong excitons, a consequence of quantum confinement and reduced screening[1–4]. Our understanding of excitonic effects in higher-order optical responses is however less complete owing to the correlated nature of the excitonic states and intricate light-matter interactions.

Second harmonic generation (SHG) is a typical nonlinear optical (NLO) response where the emitted light frequency is twice that of the incident light[5,6]. The response is characterized by a susceptibility tensor defined as the ratio of polarization density $\mathbf{P}(2\omega)$ and the light field $\mathbf{E}(\omega)$ to second order, $\chi^{\mu\nu\lambda}(2\omega;\omega,\omega) = P^{\mu}(2\omega)/(E^{\nu}(\omega)E^{\lambda}(\omega))$ where $\mu,\nu,\lambda$ are Cartesian directions. SHG spectroscopy has been widely used in characterizing the crystal structure of materials, interfaces, and strain effects owing to the sensitivity of the SHG susceptibility tensor to crystal symmetry[7–10]. Although strong SHG signals are observed in 2D materials (compared to the bulk) such as monolayer $MoS_2$ and $WSe_2$, the detailed role of excitonic effects on such enhancement is unclear[11–14].

Similarly, direct current (DC) generation from second-order optical responses (without p-n junction, called the bulk photovoltaic effect) is another topic of great fundament and practical interests such as for photovoltaic devices. Shift current is an intrinsic mechanism for the bulk photovoltaic effect and has drawn much attention through the years [15–22]. Recent work reported strong shift current in low dimensional materials, such as 2D materials[20,23] and nanotubes[19]. In particular, evidence of large excitonic effects in shift current generation in monolayer systems has been shown through direct real-time simulations of current densities that include electron-hole interactions[23].

*Ab initio* methods for calculating second-order optical responses such as SHG and shift current within the independent particle (IP) approximation are well-established[15,16,22,24]. In contrast, *ab initio* approaches including excitonic effects are still in their infancy. Based on the time-dependent perturbation theory, several studies have derived formally the so-called "sum-over-exciton-states" expressions for second order optical responses, using either a length gauge or a velocity gauge for the light-matter interaction[25–30]. In particular, the length-gauge methods[26,27] are free from unphysical low-frequency divergences[31]. However, they have not yet been formulated for practical ab initio calculations or diagrammatic interpretations. On the other hand, an *ab initio* real-time propagation of wavefunction approach has been implemented to study excitonic effects on SHG on a variety of low-dimensional materials[32,33]. An *ab initio* time-dependent adiabatic GW (TD-aGW) approach with real-time propagation of the interacting density matrix has also been developed and used to study excitonic effects on shift current and SHG[23]. Real-time propagations[23,33] can provide simultaneously information on multiple higher-order responses and at higher field intensity; however, they demand high computational costs and lack a direct picture of the detailed roles played by individual exciton states. An efficient and insightful approach for NLO responses with exciton state information from state-of-the-art GW plus Bethe-Salpeter equation (GW-BSE) calculations, is therefore highly desirable.

Motivated by these considerations, we develop in this work an *ab initio* approach based on an exciton-state coupling (ESC) formalism to study second order optical responses. We apply this approach to investigate the effects of excitons in SHG and shift current in monolayer h-BN and $MoS_2$. We show that in both materials, excitons significantly enhance the SHG spectra intensity compared to the calculations without electron-hole interactions. We identify that the large excitonic enhancement at low frequencies for monolayer h-BN is because of the unusual concurrence of bright 1s and 2p exciton states in the same material; on the other hand, in monolayer $MoS_2$, excitonic enhancement at two-phonon frequencies in resonant with the prominent A and B excitons is tiny, but it is huge with the C excitons owing to the existence of multiple bright C excitons and strong inter-exciton coupling among them. Similar behaviors are also found in shift current spectra. Our study thus explains why monolayer h-BN and $MoS_2$, despite sharing the same crystal symmetry and having low-energy band extrema at the K and K' valleys, exhibit significantly different nonlinear optical spectra.

Applying a perturbative approach to the equation of motion of the interacting density matrix[25,26], we solve the TD-aGW equation as given in Ref.[23] to different orders via a Green's function procedure in terms of the intermediate excited states of the system being exciton states. Using the length gauge for the light-matter interaction ($-e\,\mathbf{E}(t)\cdot\mathbf{r}$), to second order, we arrive at our main result for the susceptibility tensor for SHG,

$$\begin{aligned}\chi^{\mu,\nu\lambda}(2\omega\,;\omega\,,\omega\,) &= \frac{e^3}{2\epsilon_0 V}\sum_{n,m}\left[\frac{R^{\mu}_{0n}R^{\nu}_{nm}R^{\lambda}_{m0}}{(2\hbar\omega\ -E_n+i\eta)(\hbar\omega\ -E_m+i\eta)}\right.\\ &+\frac{R^{\nu}_{0n}R^{\lambda}_{nm}R^{\mu}_{m0}}{(2\hbar\omega+E_m+i\eta)(\hbar\omega+E_n+i\eta)}\\ &\left.+\frac{R^{\lambda}_{0n}R^{\mu}_{nm}R^{\nu}_{m0}}{(\hbar\omega\ -E_m+i\eta)(-\hbar\omega\ -E_n-i\eta)}\right]+(\lambda\leftrightarrow\nu\,)\end{aligned} \tag{1}$$

where $e$ is the electron charge and $V$ is the volume of the crystal. The notation $(\lambda\leftrightarrow\nu)$ means an exchange of the two Cartesian directions. $E_n$ is excitation energy of the *n*-th exciton state $|S^{(n)}\rangle$, which is expressed as $\left|S^{(n)}\right\rangle=\sum_{vc\mathbf{k}}A^{(n)}_{vc\mathbf{k}}|vc\mathbf{k}\rangle$ with $|vc\mathbf{k}\rangle$ being free electron-hole pairs of valence and conduction band states, and $A^{(n)}_{vc\mathbf{k}}$ being the **k**-space envelope function of the exciton. $R^{\nu}_{nm}$ are related to the optical coupling matrix elements between two exciton states and $R^{\nu}_{n0}$ are the ones between an exciton state and the ground state, where we have used the special notation $m = 0$ for the ground state with no excitons. It can be shown that $R^{\nu}_{n0} = \sum_{cv\mathbf{k}} A^{(n)*}_{cv\mathbf{k}} r^{\nu}_{cv\mathbf{k}}$, and $R^{\nu}_{nm}$ are derived by treating the inter-band and intra-band matrix element of the position operator $\mathbf{r}$ separately through Berry connections, which are expressed as $R^{\nu}_{nm} = Y^{\nu}_{nm} + Q^{\nu}_{nm}$ with $Y^{\nu}_{nm} = \sum_{cv\mathbf{k}} \left( \sum_{c' \neq c} A^{(n)*}_{cv\mathbf{k}} \, r^{\nu}_{cc'\mathbf{k}} \, A^{(m)}_{c'v\,\mathbf{k}} - \sum_{v' \neq v} A^{(n)*}_{cv\mathbf{k}} \, A^{(m)}_{c\,v'\,\mathbf{k}} \, r^{\nu}_{v'v\mathbf{k}} \right)$ and $Q^{\nu}_{nm} = i \sum_{cv\mathbf{k}} A^{(n)*}_{cv\mathbf{k}} \left( \partial_{k^{\nu}} A^{(m)}_{cv\mathbf{k}} - i(r^{\nu}_{cc\mathbf{k}} - r^{\nu}_{vv\mathbf{k}}) A^{(m)}_{cv\mathbf{k}} \right)$. In the above expressions, $r^{\nu}_{jj'\mathbf{k}}$ is interband (when $j \neq j'$) or intraband (when $j = j'$) Berry connection, which is given by $r^{\nu}_{jj'\mathbf{k}} = i \langle u_{j\mathbf{k}} | \partial_{k^{\nu}} | u_{j'\mathbf{k}} \rangle$ with $|u_{j\mathbf{k}}\rangle$ being the cell-periodic Bloch states. We note here that in *ab initio* calculations, the random phases associated with quantities of different wavevector **k** pose a challenge to the calculations of the **k**-derivative of the envelope function $\partial_{\mathbf{k}} A^{(n)}_{vc\mathbf{k}}$ and **k**-derivative of the wavefunctions $\partial_{\mathbf{k}} |u_{j\mathbf{k}}\rangle$. To overcome this issue, we use a locally smooth gauge construction scheme to ensure these quantities are well-defined (see Ref.[23] and Supporting Information for details). We also emphasize that Eq. 1 does not contain a divergent prefactor (like powers of $1/\omega$) and is free from numerical instabilities as $\omega \rightarrow 0$. With a similar procedure, we also derive the exciton-state coupling formalism for the shift current conductivity, as provided in Supporting Information.

The terms in Eq. 1 can be visualized using Feynman diagrams. The Feynman diagram approach for NLO responses was introduced in Refs.[34,35] at the single-particle level. Here, we extend it to describe nonlinear optics with excitonic effects. As shown in Fig. 1, the matrix elements $R_{nm}$ or $R_{n,0}$ are associated with the photon-exciton vertices (denoted by the dots) which describe a photon that couples two exciton states or connects an exciton state with the ground state. The solid lines stand for Green's functions of quantum states including exciton states and the ground state. The plotted diagram in Fig. 1 depicts the first term in Eq. 1, and by cyclic permutation of the $\{0, m, n\}$ states and exchange of the Cartesian directions $\lambda$ and $\nu$, we can obtain all terms in Eq. 1.

In the computation of $\chi^{\mu\nu\lambda}$ using the ESC formalism, the excitons' excitation energies and envelope functions in the Eq. 1 are obtained by solving the *ab initio* GW-BSE equation [36,37], as implemented in the BerkeleyGW package[38]. We have performed benchmark calculations using both the ESC formalism and the TD-aGW method for monolayer GeS (which was previously studied in Ref.[23]) as well as for monolayer h-BN and $MoS_2$, and an excellent agreement between the two methods are found (see Supporting Information).

We first study the SHG responses of monolayer h-BN, a large bandgap semiconductor with strong excitonic effects[39,40]. Figure 2(a) shows the $yyy$ component of the SHG susceptibility tensor computed at two levels of theory. In general, $\chi^{abc}$ as a tensor in 2D has 8 independent components; however, for our system with $D_{3h}$ symmetry, $\chi^{yyy} = -\chi^{yxx} = -\chi^{xxy} = -\chi^{xyx}$, with all other components equal to zero[5]. The SHG spectrum shows two sets of double peak structure, one at the energy of the 1s-like and 2p-like exciton states (denoted as peak I and II) and the other at half of their energies (denoted as $I_{1/2}$ and $II_{1/2}$). These four peaks can be understood from terms with the two-photon resonance and the single-photon resonance due to the denominators in Eq. 1. Overall, our results agree reasonably well with previous first-principles calculations using a time propagation method[33].

In the following, we investigate in more details the peaks denoted with "1/2" that are commonly focused on in SHG experiments[11,14,41]. Comparing results from the ESC formalism and the IP formalism, we observe strong excitonic enhancement to both peak $I_{1/2}$ and peak $II_{1/2}$. To understand this, we focus on the dominated term (the first term) in the square bracket of Eq. 1 and analyze the matrix elements appeared in the numerator. We define a product coupling amplitude $N_{ij} = \sum_{\{n|E_n=E_i\}} \sum_{\{m|E_m=E_j\}} R_{0,n} R_{nm} R_{m,0}$ . Here the indices $i$ and $j$ refer to the specific sets of excitons with energy $E_i$ and $E_j$, which can have multiple degenerate states, and all the degenerate exciton states associated with these energies are included in the calculation of $N_{ij}$. We have dropped the Cartesian direction $y$ in $R_{nm}^{y}$ and $R_{m,0}^{y}$ for notational simplicity since we are concerned with the yyy component. The first term of the summation in Eq. 1 now can be rewritten as $\sum_{ij} N_{ij} (2\hbar\omega - E_i + i\eta)^{-1} \left(\hbar\omega - E_j + i\eta\right)^{-1}$. Due to the presence of the denominator $2\hbar\omega - E_i + i\eta$, there would be a large SHG intensity when $\omega$ is near half of exciton energy $E_i$, as long as there exist large coupling amplitudes $N_{ij}$ in the set $\{N_{ij}, j = \text{all}\}$. This argument is visualized in Fig. 2(b). In the figure, the absolute values of $N_{ij}$ are represented by a series of dots with different radii according to their amplitude. The lower orange bracket in the figure indicates that peak $I_{1/2}$ is mainly related to the set of coupling amplitudes $\left\{N_{ij} | i = 1s, j = \text{all}\right\}$. Within this set, $N_{i=1s,j=2p}$ and $N_{i=1s,j=1s}$ dominate and they are the main source for the large intensity of peak $I_{1/2}$. Similarly, peak $II_{1/2}$ is related to the set $\left\{N_{ij} | i = 2p, j = \text{all}\right\}$, which is indicated by the upper orange bracket and dominated by two different couplings.

Let's focus on the largest coupling amplitude $N_{i=1s,j=2p}$ indicated by the red arrow in Fig. 2(b) to get some physical insight of the excitonic effect in the SHG process for peak $I_{1/2}$. This coupling amplitude involves two degenerate 1s-like exciton states (one from the K, the other from the K' valley) and two degenerate *optically bright* 2p-like states (also one from the K, the other from the K' valley). Interestingly, in monolayer h-BN, the oscillator strength for excitation from the ground state to the 2p-like states (i.e., $R_{2\text{p},0}$) is large and comparable to that of excitation to the 1s-like states (i.e., $R_{1\text{s},0}$), with $\left|R_{2p,0}\right| \approx 0.5 \left|R_{1s,0}\right|$. This unusual brightness for the 2p excitons in dipole-allowed interband transition systems is attributed to a large trigonal wrapping effect[40,42,43]. This effect is reflected in the **k**-space envelope functions of 2p-like excitons in monolayer h-BN, which are significantly distorted from a circular shape (Fig. 2(d)). Moreover, the bright 1s-like states and 2p-like states from the same valley can be coupled by the **r** operator, since their angular momenta differ by 1. As a result, all the coupling matrix elements (vertices) in Fig. 2(c) are large, leading to a substantial coupling amplitude $N_{i=1s,j=2p}$ and thus the significant excitonic enhancement seen in peak $I_{1/2}$. The similar mechanism is also the origin of the large excitonic enhancement for peak $II_{1/2}$.

The strong excitonic enhancement in SHG in the low frequency regime for monolayer h-BN needs not be a general feature for other materials. It is well-known that peak A and peak B in the monolayer $MoS_2$'s linear absorption spectrum originate from the 1s excitons of the A and B series at the $K$ and $K'$ valley. Their difference in energy mainly corresponds to the splitting of the top of the valence band at the K/K' point by spin-orbit coupling. At frequencies above the quasiparticle band gap, peak C (which consists of correlated electron-hole pairs near $\Gamma$, $K$ and $K'$ valleys) is the most pronounced peak[1]. Both the A/B and C excitons feature large linear optical transition matrix elements due to strong excitonic effects (Fig. 3a). Therefore, one would expect similar exciton enhancement for both the A/B and C "1/2" peaks in the SHG intensity; however, this is not the case for the following reasons from our results.

In Fig. 3(b), we show the computed yyy component of the SHG susceptibility tensor of monolayer $MoS_2$. We label the peaks at 0.97 eV, 1.05 eV and 1.35 eV from the ESC results as peak $A_{1/2}$, $B_{1/2}$ and $C_{1/2}$ since they are at half of the energy of peak A, peak B and peak C in the linear absorption spectrum. We find that the peak $A_{1/2}$ / $B_{1/2}$ intensity is close to the value of the low-frequency SHG intensity from the IP calculation, while there is a three-fold excitonic enhancement in the intensity of peak $C_{1/2}$ compared to the IP peak intensity at the corresponding interband transition energies. The dominance of peak $C_{1/2}$ agrees with experimental findings[12] (see Fig. 4b) as well as calculations based on real-time propagation studies[33,44] and tight-binding model results[45] although no deep understanding was provided in these previous studies.

To understand the distinctively different enhancement effects on peak $C_{1/2}$ and peak $A_{1/2}$ / $B_{1/2}$, we plot the product coupling amplitude $N_{ij}$ for monolayer $MoS_2$ in Fig. 3(c). We first look at peak $C_{1/2}$. It is mostly related to the set of coupling amplitudes depicted by the upper orange brackets in Fig. 3(c), since they will contribute most when the two-photon energy is at resonant with the energy of the C exciton states. We find that many coupling amplitudes $N_{ij}$ in this set exhibit large values. This is understood as follows. Two bright excitons (say $|C_m\rangle, |C_n\rangle$) in the series C can coupled strongly via the **r** operator because their envelope functions exhibit a large degree of trigonal warping and are distributed in a similar region in reciprocal space (see Supporting Information). As a result, the three coupling elements $(R_{0,C_n}, R_{C_nC_m}, R_{C_m,0})$ can be simultaneously large (as indicated in the diagram in Fig. 3(d)), leading to a substantial product coupling amplitude. Due to the presence of multiple bright C exciton states, there are many combinations for large-valued coupling amplitudes, which together result in a giant excitonic enhancement for the peak $C_{1/2}$.

On the other hand, the small excitonic enhancement at peak $A_{1/2}$ can be understood by analyzing the product coupling amplitude within the set denoted by the lower orange brackets in Fig. 3(c) (The behavior of peak $B_{1/2}$ can be understood in a similar way). Among this set, $N_{i=1s,j=2p}$ is the dominant one, denoted by the green arrow in Fig. 3(c). However, its magnitude is one order of magnitude smaller than some of those in the C series. This is because one of its constituent elements $R_{2p,0}$, which corresponds to the optical oscillator strength of the 2p exciton, is small, with only 0.06 times the magnitude of $R_{1s,0}$, as indicated in Fig. 3(e). The smallness of $R_{2p,0}$ is related to the small degree of trigonal warping of the exciton wavefunction in the K/K' valleys of monolayer $MoS_2$[42,43]. Besides $N_{i=1s,j=2p}$, the other coupling amplitude are all small, since there are no other bright excitons that couple strongly with the 1s excitons.

Finally, we show that our approach to understanding excitonic enhancement associated with different exciton states in SHG can be applied to understanding shift current generation. In Fig. 4a, we plot the yyy component of shift current conductivity tensor for monolayer $MoS_2$. The A and B peaks are around 40 times smaller than that of the C peak. The physical picture for this is similar to that in SHG, since the optical processes involved in shift current generation also include the product of three coupling terms connecting the ground state and two intermediate exciton states (see Fig. S1 and Eq. S1 in the Supporting Information). Due to the lack of intermediate bright excitons for the A (or B) exciton to couple with, its resonance in the shift current generation process is weak, resulting in very low intensity. In contrast, monolayer h-BN behaves oppositely because of the coexistence of bright 1s and bright 2p excitons, exhibiting significant excitonic enhancement of shift current in its low-energy peaks (see Supporting Information for more details), which is similar to the SHG case.

In conclusion, we have developed an efficient method based on an exciton-state coupling formulation making use of Berry connections in the length gauge to compute nonlinear optical responses with excitonic effects from first principles. Applying this method to monolayer h-BN and $MoS_2$, we have elucidated the microscopic origin of excitonic enhancements on their SHG and shift current responses. A comparison of the two materials suggests strong trigonal warping is essential for large excitonic enhancement in this class of hexagonal 2D materials[13]. The exciton-state coupling analysis developed in this work can be applied to understanding the excitonic effect in other nonlinear optical phenomena, such as difference frequency generation and sum frequency generation[12].

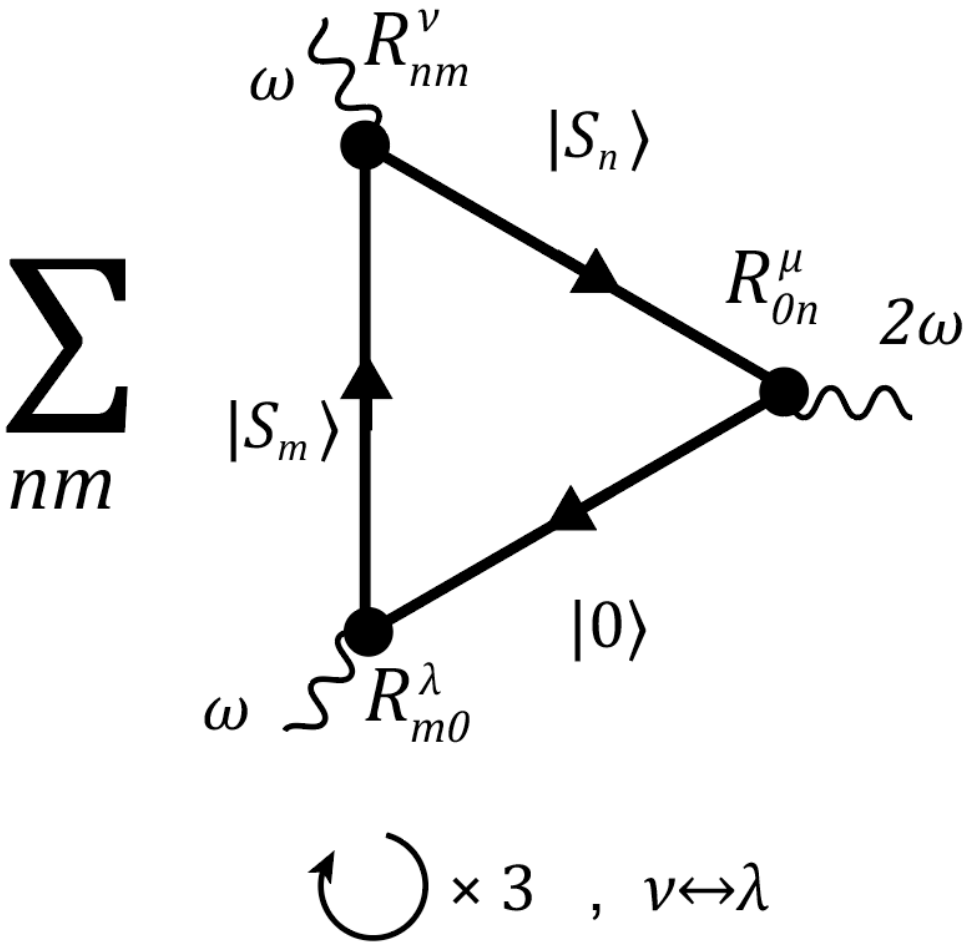

**Figure 1.** Feynman diagrams for second harmonic generation (SHG) in a framework of coupling of exciton states. The solid lines stand for Green's functions of different states, and the wavy lines refer to external photons. $\omega$ represents the incoming frequency and $2\omega$ represents the outgoing frequency. $R^{\nu}_{nm}$ and $R^{\nu}_{n,0}$ denote matrix elements coupling different exciton states and are associated with the photon-exciton vertices (denoted by the dots). $\lambda, \nu, \mu$ are Cartesian directions of the electric field of light. The symbol $\circlearrowright \times 3$ represents a cyclic permutation of the $\{0, m, n\}$ labels and the symbol $\lambda \leftrightarrow \nu$ represents an exchange of the two Cartesian directions. In total, there are six distinct diagrams, and the sum is over all exciton states with indices $n$ and $m$.

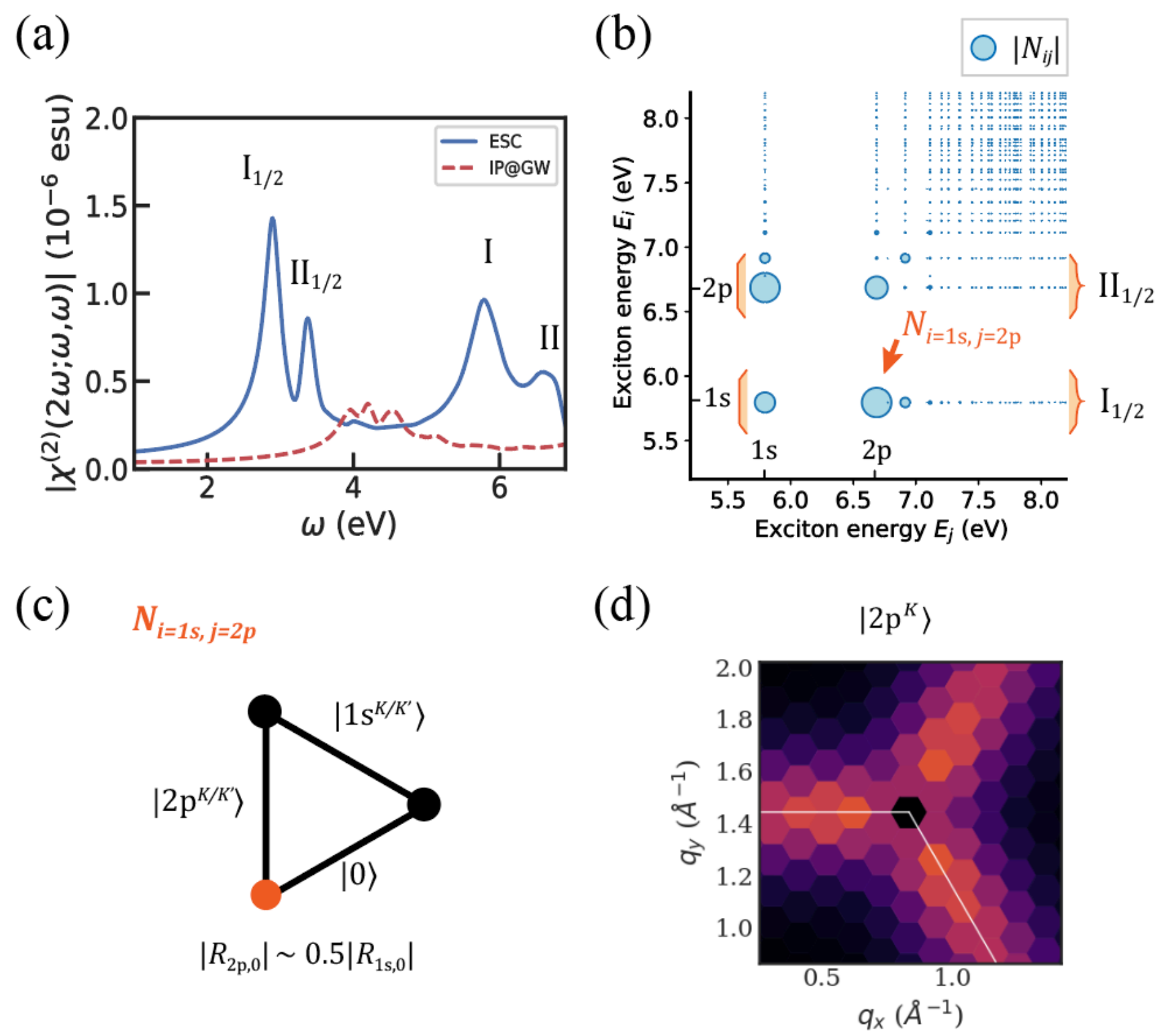

**Figure 2.** (a) Absolute value of the yyy component of the SHG susceptibility spectrum for monolayer h-BN. Blue solid line is the result from Eq. (1), the exciton-state coupling (ESC) formalism; red dashed line is the result from the independent particle (IP) formalism using GW quasiparticle energies. (b) Modules of coupling amplitudes $|N_{ij}|$. The magnitude of $|N_{ij}|$ is proportional to the radius of the dot. Orange brackets are used to outline the groups of $N_{ij}$ that contribute to the main peaks $\mathrm{I}_{1/2}$ and $\mathrm{II}_{1/2}$ in the spectrum. (c) The diagram corresponding to $N_{i=1s,j=2p}$ denoted by the red arrow in (b). The red dot emphasizes that the matrix element of $R_{2p,0}$ is unusually large for materials with dipole allowed interband transitions. (d) **k**-space exciton envelope function of the $|2p^K\rangle$ exciton state in monolayer h-BN.

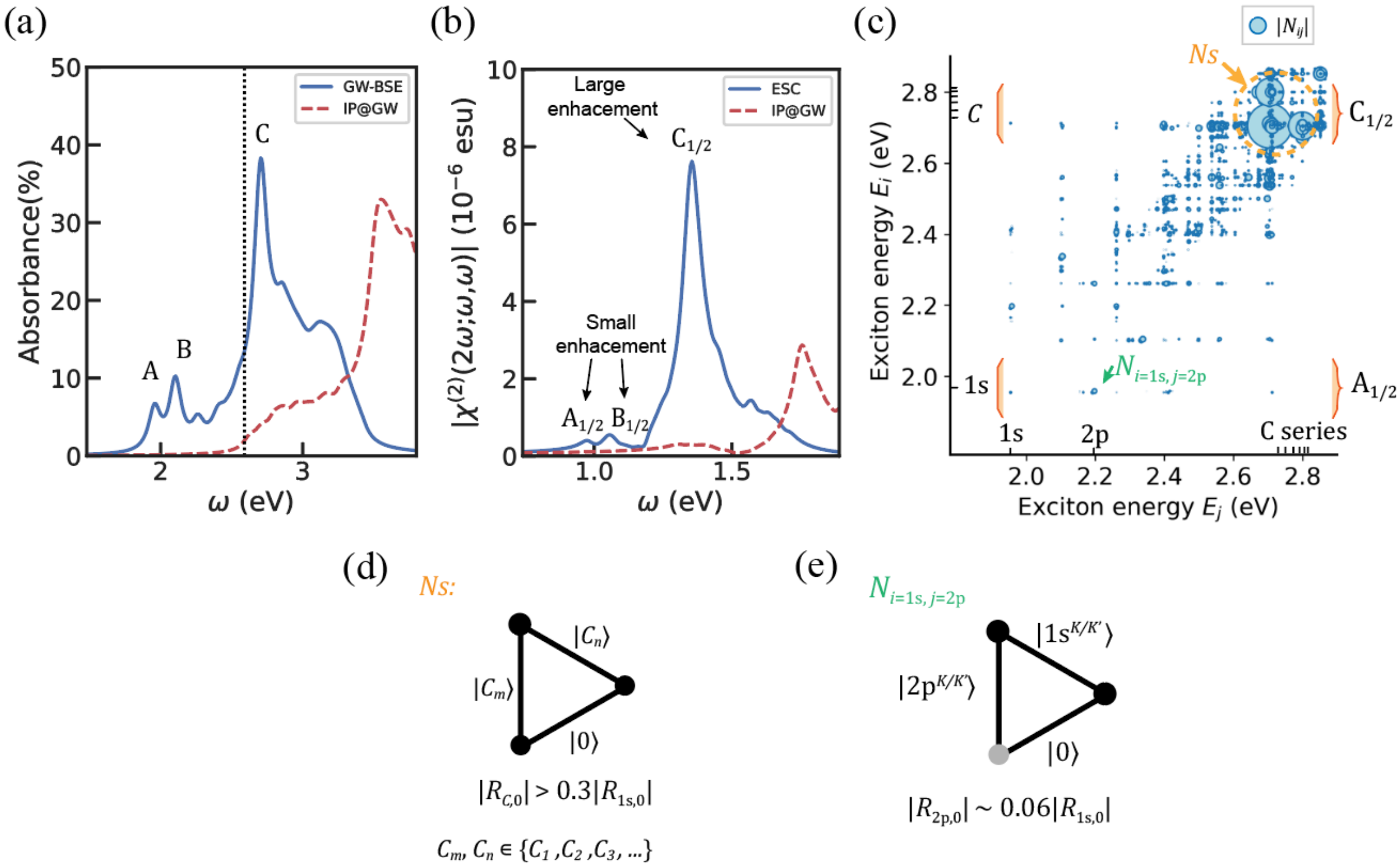

**Figure 3.** (a) Absorbance spectrum of monolayer $MoS_2$. The dotted line shows the energy of GW bandgap. (b) Absolute value of the yyy component of the SHG susceptibility spectrum for monolayer $MoS_2$. Blue solid line is the result from the ESC formalism; red dashed line is the result from the IP formalism using GW quasiparticle energies. A broadening parameter of $\eta = 50$ meV is used in both (a) and (b). (c) Modules of coupling amplitudes $|N_{ij}|$. Orange brackets are used to outline the groups of $N_{ij}$ that contribute dominantly to the $A_{1/2}$ and $C_{1/2}$ peaks in the spectrum. The groups of $N_{ij}$ related to the $B_{1/2}$ peak is not explicitly labeled for notational simplicity since the analysis is similar to that of $A_{1/2}$ peak. (d) The diagram corresponding to the group of coupling amplitudes which are enclosed by the dashed orange circle in (c). (e) The diagram corresponding to $N_{i=1s,j=2p}$ denoted by the green arrow in (c). The gray dot emphasizes that the matrix element is small.

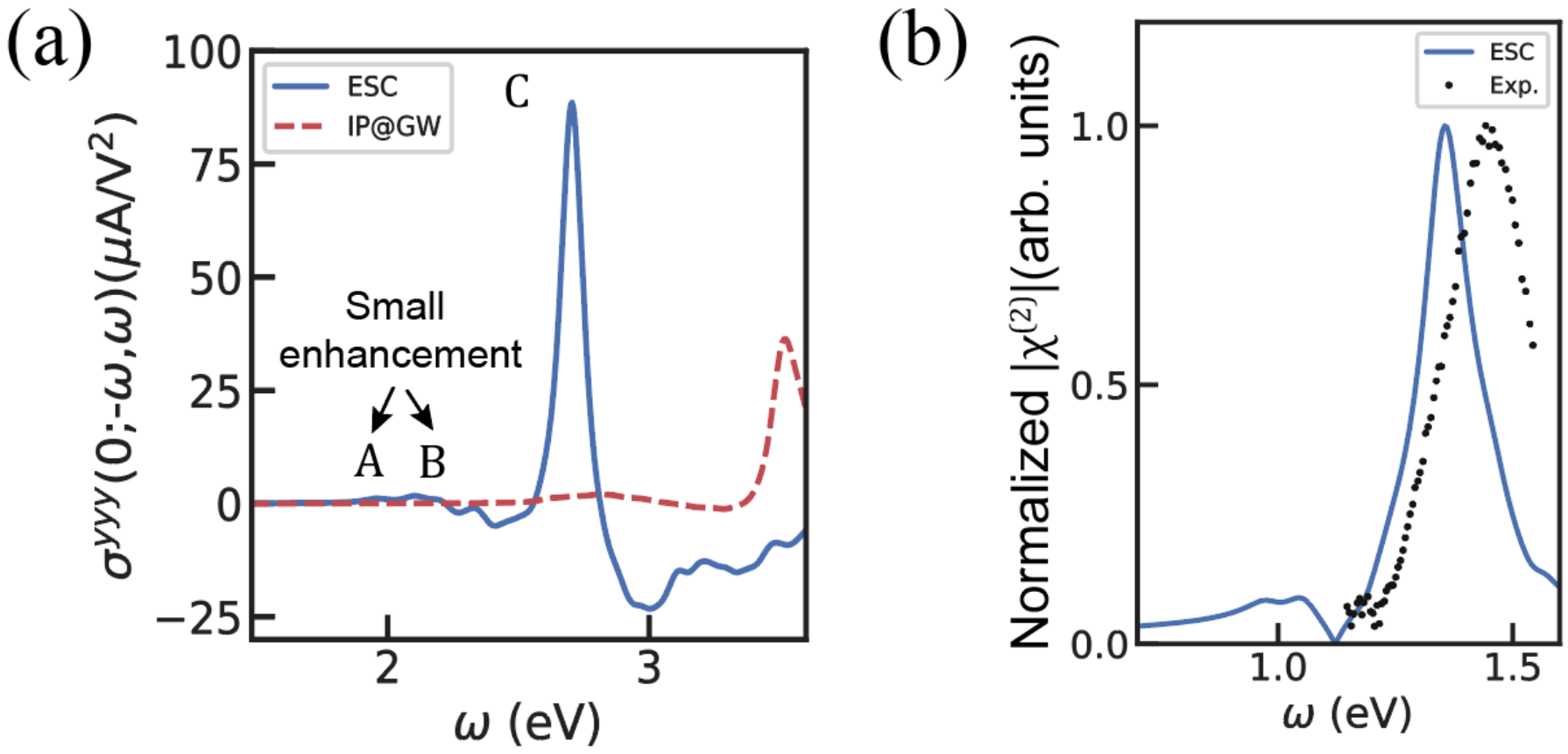

**Figure 4.** (a) Frequency dependence of the yyy-component of the shift current conductivity tensor of monolayer $MoS_2$ computed with different approaches. A broadening parameter of $\eta = 50$ meV is used. (b) Comparison of the normalized absolute value of SHG susceptibility of monolayer $MoS_2$ from experiment[12] with that from our theoretical calculations using exciton-state coupling formalism. The intensity is normalized to its maximum value for each dataset. A broadening parameter of $\eta = 80$ meV is used in the theoretical spectrum for a better comparison with experiment. More details about this comparison can be found in the Supporting Information.

**Corresponding Authors**

* yanghao@gate.sinica.edu.tw

* sglouie@berkeley.edu

**Notes**

The authors declare no competing financial interest.

**Supporting Information**

Computational details of GW-BSE, formalism for shift current conductivity, locally smooth gauge construction scheme, benchmark calculations, comparison of SHG spectrum with experiment data, absorbance spectrum, exciton envelope functions.

**Acknowledgements**

This work is primarily supported by the National Science Foundation under grant DMR-2325410 which provided the exciton-state coupling density matrix perturbation theory formulation, Feynman diagram studies, symmetry analyses of exciton couplings, and the *ab* *initio* second harmonic generation calculations. It is partially supported by the Center for Computational Study of Excited State Phenomena in Energy Materials (C2SEPEM) funded by the U.S. Department of Energy, Office of Science, Basic Energy Sciences, Materials Sciences and Engineering Division under Contract No. DE-AC02-05CH11231, as part of the Computational Materials Sciences Program which provided GW-BSE calculations, exciton-basis coupling matrix elements, and advanced codes. We acknowledge the National Energy Research Scientific Computing Center (NERSC) which is a DOE Office of Science User Facility supported by the Office of Science of the U.S. Department of Energy under Contract No. DE-AC02-05CH11231, the Texas Advanced Computing Center (TACC) which is a NSF User Facility supported by National Science Foundation under Grant No. OAC-1818253, and the National Center for High-performance Computing (NCHC) in Taiwan for providing high performance computing resources that have contributed to the research results reported within this paper.

**References**

1. Qiu, D. Y., Felipe, H. & Louie, S. G. Optical spectrum of MoS 2: many-body effects and diversity of exciton states. *Physical Review Letters* **111**, 216805 (2013).

2. Qiu, D. Y., Felipe, H. & Louie, S. G. Screening and many-body effects in two-dimensional crystals: Monolayer MoS 2. *Physical Review B* **93**, 235435 (2016).

3. Chernikov, A. *et al.* Exciton Binding Energy and Nonhydrogenic Rydberg Series in Monolayer WS2. *Physical Review Letters* **113**, 076802 (2014).

4. Wang, G. *et al.* Colloquium: Excitons in atomically thin transition metal dichalcogenides. *Reviews of Modern Physics* **90**, 21001 (2018).

5. Boyd, R. W. Nonlinear optics. in *Springer Handbook of Atomic, Molecular, and Optical Physics* 1097–1110 (Springer, 2008).

6. Luppi, E. & Véniard, V. A review of recent theoretical studies in nonlinear crystals: towards the design of new materials. *Semiconductor Science and Technology* **31**, 123002 (2016).

7. Kumar, N. *et al.* Second harmonic microscopy of monolayer MoS_2. *Phys. Rev. B* **87**, 161403 (2013).

8. Liang, J. *et al.* Monitoring Local Strain Vector in Atomic-Layered MoSe2 by Second-Harmonic Generation. *Nano Letters* **17**, 7539–7543 (2017).

9. Mennel, L. *et al.* Optical imaging of strain in two-dimensional crystals. *Nature Communications* **9**, (2018).

10. Li, Y. *et al.* Probing symmetry properties of few-layer MoS2 and h-BN by optical second-harmonic generation. *Nano letters* **13**, 3329–3333 (2013).

11. Malard, L. M., Alencar, T. V., Barboza, A. P. M., Mak, K. F. & de Paula, A. M. Observation of intense second harmonic generation from MoS_2 atomic crystals. *Phys. Rev. B* **87**, 201401 (2013).

12. Yao, K. *et al.* Continuous wave sum frequency generation and imaging of monolayer and heterobilayer two-dimensional semiconductors. *ACS nano* **14**, 708–714 (2019).

13. Säynätjoki, A. *et al.* Ultra-strong nonlinear optical processes and trigonal warping in MoS2 layers. *Nature Communications* **8**, 893 (2017).

14. Kikuchi, Y. *et al.* Multiple-peak resonance of optical second harmonic generation arising from band nesting in monolayer transition metal dichalcogenides TX2 on SiO2/Si(001) substrates (T=Mo,W;X=S,Se). *Physical Review B* **100**, 075301 (2019).

15. Aversa, C. & Sipe, J. E. Nonlinear optical susceptibilities of semiconductors: Results with a length-gauge analysis. *Physical Review B* **52**, 14636–14645 (1995).

16. Young, S. M. & Rappe, A. M. First Principles Calculation of the Shift Current Photovoltaic Effect in Ferroelectrics. *Physical Review Letters* **109**, 116601 (2012).

17. Nakamura, M. *et al.* Shift current photovoltaic effect in a ferroelectric charge-transfer complex. *Nature Communications* **8**, 281 (2017).

18. Osterhoudt, G. B. *et al.* Colossal mid-infrared bulk photovoltaic effect in a type-I Weyl semimetal. *Nature Materials* **18**, 471–475 (2019).

19. Zhang, Y. J. *et al.* Enhanced intrinsic photovoltaic effect in tungsten disulfide nanotubes. *Nature* **570**, 349–353 (2019).

20. Akamatsu, T. *et al.* A van der Waals interface that creates in-plane polarization and a spontaneous photovoltaic effect. *Science* **372**, 68–72 (2021).

21. Sotome, M. *et al.* Terahertz emission spectroscopy of ultrafast exciton shift current in the noncentrosymmetric semiconductor CdS. *Phys. Rev. B* **103**, L241111 (2021).

22. Sipe, J. E. & Shkrebtii, A. I. Second-order optical response in semiconductors. *Physical Review B* **61**, 5337–5352 (2000).

23. Chan, Y.-H., Qiu, D. Y., Jornada, F. H. da & Louie, S. G. Giant exciton-enhanced shift currents and direct current conduction with subbandgap photo excitations produced by many-electron interactions. *Proceedings of the National Academy of Sciences* **118**, e1906938118 (2021).

24. Luppi, E., Hübener, H. & Véniard, V. Ab initio second-order nonlinear optics in solids: Second-harmonic generation spectroscopy from time-dependent density-functional theory. *Phys. Rev. B* **82**, 235201 (2010).

25. Pedersen, T. G. Intraband effects in excitonic second-harmonic generation. *Phys. Rev. B* **92**, 235432 (2015).

26. Taghizadeh, A. & Pedersen, T. G. Gauge invariance of excitonic linear and nonlinear optical response. *Phys. Rev. B* **97**, 205432 (2018).

27. Mkrtchian, G. F., Knorr, A. & Selig, M. Theory of second-order excitonic nonlinearities in transition metal dichalcogenides. *Phys. Rev. B* **100**, 125401 (2019).

28. Chang, E. K., Shirley, E. L. & Levine, Z. H. Excitonic effects on optical second-harmonic polarizabilities of semiconductors. *Physical Review B* **65**, 035205 (2001).

29. Leitsmann, R., Schmidt, W., Hahn, P. & Bechstedt, F. Second-harmonic polarizability including electron-hole attraction from band-structure theory. *Physical Review B* **71**, 195209 (2005).

30. Xuan, F., Lai, M., Wu, Y. & Quek, S. Y. Exciton-enhanced spontaneous parametric down-conversion in two-dimensional crystals. *Physical Review Letters* **132**, 246902 (2024).

31. Ma, Q., Grushin, A. G. & Burch, K. S. Topology and geometry under the nonlinear electromagnetic spotlight. *Nature Materials* **20**, 1601–1614 (2021).

32. Attaccalite, C. & Grüning, M. Nonlinear optics from an ab initio approach by means of the dynamical Berry phase: Application to second- and third-harmonic generation in semiconductors. *Phys. Rev. B* **88**, 235113 (2013).

33. Grüning, M. & Attaccalite, C. Second harmonic generation in h-BN and MoS_2 monolayers: Role of electron-hole interaction. *Phys. Rev. B* **89**, 081102 (2014).

34. Parker, D. E., Morimoto, T., Orenstein, J. & Moore, J. E. Diagrammatic approach to nonlinear optical response with application to Weyl semimetals. *Physical Review B* **99**, 045121 (2019).

35. Noblet, T., Busson, B. & Humbert, C. Diagrammatic theory of linear and nonlinear optics for composite systems. *Phys. Rev. A* **104**, 063504 (2021).

36. Hybertsen, M. S. & Louie, S. G. Electron correlation in semiconductors and insulators: Band gaps and quasiparticle energies. *Physical Review B* **34**, 5390 (1986).

37. Rohlfing, M. & Louie, S. G. Electron-hole excitations and optical spectra from first principles. *Physical Review B* **62**, 4927 (2000).

38. Deslippe, J. *et al.* BerkeleyGW: A massively parallel computer package for the calculation of the quasiparticle and optical properties of materials and nanostructures. *Computer Physics Communications* **183**, 1269–1289 (2012).

39. Galvani, T. *et al.* Excitons in boron nitride single layer. *Phys. Rev. B* **94**, 125303 (2016).

40. Zhang, F. *et al.* Intervalley Excitonic Hybridization, Optical Selection Rules, and Imperfect Circular Dichroism in Monolayer h-BN. *Phys. Rev. Lett.* **128**, 047402 (2022).

41. Wang, G. *et al.* Giant Enhancement of the Optical Second-Harmonic Emission of WSe_2 Monolayers by Laser Excitation at Exciton Resonances. *Phys. Rev. Lett.* **114**, 097403 (2015).

42. Cao, T., Wu, M. & Louie, S. G. Unifying optical selection rules for excitons in two dimensions: Band topology and winding numbers. *Physical Review Letters* **120**, 087402 (2018).

43. Zhang, X., Shan, W.-Y. & Xiao, D. Optical selection rule of excitons in gapped chiral fermion systems. *Physical Review Letters* **120**, 077401 (2018).

44. Beach, K., Lucking, M. C. & Terrones, H. Strain dependence of second harmonic generation in transition metal dichalcogenide monolayers and the fine structure of the C exciton. *Phys. Rev. B* **101**, 155431 (2020).

45. Trolle, M. L., Seifert, G. & Pedersen, T. G. Theory of excitonic second-harmonic generation in monolayer MoS_2. *Phys. Rev. B* **89**, 235410 (2014).

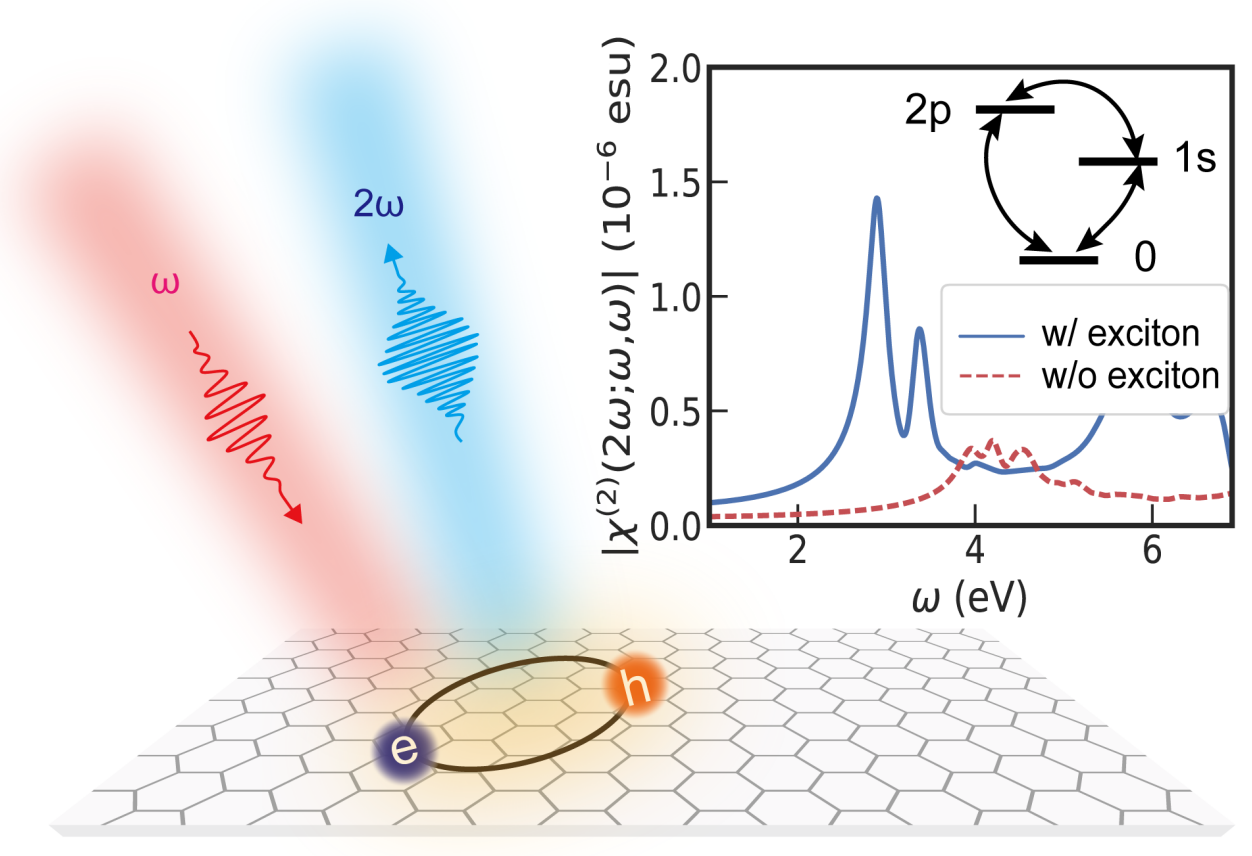

TOC Graphic